\documentclass[doublespacing]{elsart}
\usepackage{amssymb}
\begin{document}

\begin{frontmatter}
\title{ Structure and distribution of arches in shaken hard sphere deposits.}
\author[label1]{ Luis A. Pugnaloni\corauthref{a}}
\ead{l.a.pugnaloni@food.leeds.ac.uk}
\author[label2]{ G. C. Barker}
\corauth[a]{Author to whom correspondence should be addressed. Tel:+44 113 3432979 Fax:+44 113 3432982}
\address[label1]{Procter Department of Food Science, University of Leeds, Leeds LS2 9JT, UK}
\address[label2]{ Institute of Food Research, Norwich Research Park, Conley, Norwich NR4 7UA, United Kingdom.}

\begin{abstract}
\smallskip
We investigate the structure and distribution of arches formed by spherical, hard particles shaken in an external field after they come to rest. Arches (or bridges) are formed during a computer-simulated, non-sequential deposition of the spheres after each shaking cycle. We identify these arches by means of a connectivity criterion and study their structural characteristics and spatial distribution. We find that neither the size distribution nor the shape of the arches is strongly affected by the packing fraction of the deposit. Conversely, the spatial distribution and orientation of the bridges do depend on the volume fraction occupied by the spheres.
\end{abstract}


\end{frontmatter}
 \section{Introduction}

The structure of hard sphere systems has been investigated as a model for liquids (see for example Ref. \cite{Hansen}) and also as a model for granular materials \cite{Mehta,Tassopoulos}. However, the term ``hard spheres" does not define the model completely. For model liquids, we normally assume that the spheres interact through elastic collisions, whereas granular materials are modelled as inelastic, and often rough, hard spheres. Due to energy dissipation, granular systems come to rest soon after any initial velocities are imposed to the grains, and so a mechanism to excite the system, such as shaking or stirring, is necessary in order to explore new configurations. Of course, this type of mechanism is not necessary for an ensemble of elastic spheres to explore the phase space. However, if no external forces are applied to the system, the structures obtained in both cases are very similar for a given volume fraction.

A completely different situation arises when an external force acts on the spheres. Whereas the elastic spheres will flow smoothly, under an external force, inelastic spheres will tend to jam at sufficiently high volume fractions. Elastic hard spheres poured in a container under the action of gravity will fill the vessel in a liquid-like fashion (the hydrostatic pressure decreasing as we move upwards from the bottom of the vessel). On the other hand, inelastic spheres, after all the kinetic energy had been dissipated, will rest on top of each other (the forces being transmitted by the contact network rather than collisions). Despite this difference in behaviour, some structural indicators such as the pair correlation function of the system reveal no essential difference between elastic and inelastic hard spheres \cite{Mehta}.

One of the most distinctive features in the structure of a bed of inelastic hard spheres is the presence of arches. Arches (also called arcs or bridges) are multi-particle structures. The particles in an arch are said to be mutually stable (see Fig. 1 and 2). That is to say, if any particle of the arch were removed, the entire set of particles belonging to the arch would collapse under the effect of gravity. This type of structure can only be present in a system that is non-sequentially deposited, since two particles have to meet at the same time they reach the stable positions in order to achieve a mutually stable configuration.

Arches are the basic building blocks for the structural properties of granular materials. They are responsible for the non-uniform stress propagation \cite{OHern}, for the blockage of outlets during forced flow \cite{To,Zuriguel}, for the voids that determine the volume fraction \cite{Nolan,Nowak}, etc.
        
In a previous work \cite{Pugnaloni}, we have identified the arches present in computer-generated, hard sphere deposits and analysed their size distribution and general shape characteristics. The main findings were: a) about 80 per cent of the arches are string-like arches, b) the size distribution of the bridges does not strongly depend on the packing fraction of the system, c) the string-like bridges have a ``super-diffusive" like structure, and d) the distribution of the lateral extension of the bridges resembles the distribution of the normal forces found experimentally \cite{OHern} in dense packings of hard particles.

All the arch properties analysed in our previous work showed no clear dependence on the actual volume fraction of the packing of spheres. In this paper, we present new results which show that descriptors such as the orientation and spatial distribution of the arches depend strongly on the volume fraction of the system. In addition, we present a detailed description of the technique we use to identify arches in a computer generated granular system, and a discussion about the relationship between the mean coordination number and the mutual stability of bridge-particles. 

\section{\label{simul}Simulation technique }

We have examined bridge structures in hard assemblies that are generated by an established, non-sequential, restructuring algorithm \cite{Mehta2,Barker}. This algorithm restructures a stable, hard sphere deposit in three distinct stages. Initially, free volume is introduced homogeneously throughout the system and the particles are given small, random, displacements. Then, the packing is compressed in a uniaxial external field using a low-temperature Monte Carlo process. Finally, the spheres are stabilized using a steepest descendent ``drop and roll" dynamics to find a local minimum of the potential energy. Crucially, during the last part of the restructuring process, the spheres, although moved in sequence, are able to roll in contact with spheres that are in either stable or unstable positions. As a consequence, mutual stabilization may arise. The final configuration has a well-defined network of contacts and each sphere is supported, in its locally stable position, through point contacts, by a set of three other spheres uniquely defined. In practice, the final configuration may include a few non-stabilized particles.

The simulations are performed in a squared prism (base size: $6 \sigma \times 6 \sigma$, where $\sigma$ is the diameter of the spheres). Periodic boundary conditions are applied in the $x$- and $y$-direction, and a hard disordered base at $z=0$ limits the ``fall" of the particles as they are subjected to the external field applied downwards in the $z$-direction. Our previous investigations \cite{Mehta2} have shown that this restructuring process does not depend strongly on the simulation parameters and that, after many cycles, restructured packings reach a steady state in the mean packing fraction and the mean coordination number. The nature of the steady state is dictated by the amount of free volume introduced in the expansion phase of the restructuring process (the shaking amplitude) \cite{Mehta2}. We have shown \cite{Barker,Barker2} that the random packings generated in this way have many features in common with the states generated in vibrated granular media. In particular, we have shown that we can explore ``irreversible" and ``reversible" branches of the density versus a systematically varying shaking amplitude, in agreement with the experimental findings \cite{Nowak}.

The system is mono-disperse in order to avoid segregation effects. Additionally the disordered base prevents the system from ordering. The total number of particles (spheres) introduced in the simulation box is $N_{tot}=2200$. We start with a sequential configuration where all the spheres were deposited one at a time in the container (from an infinitely high initial position) so that no mutual stabilization is present. Then, we apply a large number of restructuring cycles to reach the steady state for a given shaking amplitude. Finally, about 100 stable configurations (picked every 100 cycles in order to avoid correlation effects) are saved for future analysis.
We use three different shaking amplitudes, which gives us a set of three different packings. The final average volume fractions for these systems are $\phi=0.560$, $0.580$ and $0.595$. The coordination number is in all three cases $Z \approx 4.6 \pm 0.1$.

\section{ Arch definition }

An arch or bridge is a multi-particle structure characterized by the mutual stabilization of its elements. When two or more particles meet during the deposition process in a granular material they can reach a mutually stable configuration in which every particle in the set is essential for the stability of the others.

In Fig. 1 we show examples of two-dimensional bridges. Every bridge in two dimensions is a chain of particles supported by two base-particles, which we do not consider as part of the bridge itself, at its ends. Each particle in the bridge (and in the packing in general) is supported by two other particles. If the bridge is of size one ---we call this a one-particle bridge or a sequentially-deposited particle--- the only particle in the bridge is supported by the two base-particles. The two base-particles do not need the bridge-particle to stabilize them; they are stabilized by other particles in the assembly. For multi-particle bridges, which are the actual non-sequential structures in the packing, each bridge-particle supports and is supported by at least another bridge-particle. If a bridge-particle is in contact with a base-particle, then the former is supported by the base and a second bridge-particle while it contributes to the stability of this second particle of the bridge. If a particle is not in contact with a base-particle, then it is supported by two other bridge-particles and it contributes to the stabilization of these two. Therefore, a bridge is stable thanks to the contributions of every particle in it.

In three dimensions, a particle stabilized by point contacts needs three supporting points (see Fig. 2). Then, a particle that belongs to a bridge in three dimensions can be stabilized by: a) three base-particles (only in one-particle bridges), b) two base-particles and a second bridge-particle, c) one base-particle and two extra bridge-particles, or d) three extra bridge-particles. An immediate consequence of this is that three-dimensional arches can have several branches and are not single chains as they are in two dimensions. However, the same mutual stability concept applies to define which particles belong to a given arch and which particles form the base of the bridge.

To define a bridge (or arch) rigorously, we use a connectivity criterion. We say that two grains in a packing are connected if they are mutually stable, i.e. they contribute to the stability of each other. Then a bridge is a set of particles such that we can trace a path of connected particles between any pair in the set. This definition guaranties that in any packing every particle belongs to a bridge (we include one-particle bridges as an special case) and that there is not intersection between bridges, i.e. no particle belong to two or more bridges. It is worth mentioning that this definition is entirely analogous to the definition of clusters used in the study of aggregation and percolation \cite{Stauffer}.
 In addition to the definition of a bridge, we can define the base of a bridge as the set of particles that contribute to the stability of at least one particle of a given bridge but is not supported by any particle of that bridge. In Fig. 2 we show three-dimensional bridges extracted from our simulations, by using the above definition, where all the remaining particles of the system were removed for clarity except for the base particles which are displayed with a different colour.

In practice, bridges can be identified by making a list of mutually stabilized pairs of particles and then using any standard cluster counting algorithm \cite{Stoddard,Stauffer} to separate the system into disjoint sets of connected particles.

\section{Structural descriptors}

Once the bridges in a given particle deposit have been identified, different descriptors can be used to analyse their structure and distribution. In a previous work \cite{Pugnaloni}, we showed that some descriptors are not sensitive to the volume fraction of the deposit. In this paper, we expand our search for descriptors that do show a dependence on the volume fraction of the assembly of spheres. In this section we define several descriptors for the bridges and apply these definitions to analyse the bridges encountered in the three packings generated with our restructuring algorithm (see section \ref{simul}).

\subsection{Coordination number}

We start by indicating the close relationship between the state of mutual stability of the spheres and the mean coordination number of the packing. The state of mutual stability $\alpha$ of a sphere is the number of supporting point contacts that are mutually stabilizing contacts. The possible values for $\alpha$ are 0, 1, 2 and 3 corresponding respectively to the stability situations (a), (b), (c) and (d) described in the previous section. In Fig. 3 we present the histograms for the probability $p(\alpha)$ that a given particle in the system is in state $\alpha$. We show results for the three non-sequentially deposited systems generated as we described in section \ref{simul}. As we can see, non-sequentially deposited spheres are mostly mutually stable with one or two contacting neighbours.

Each particle contributes differently to the total number of point contacts in the packing depending on its state $\alpha$. If the spheres are deposited sequentially, the total number of point contacts in the deposit can be obtained by counting the number of {\it supporting} point contacts for each sphere ---which is three--- and multiplying by the number of spheres in the packing. However, if two spheres stabilize each other, the same {\it supporting} point contact would be counted twice. Therefore, in a non-sequentially reorganized deposit, the non-mutually stabilizing supporting point contacts of a given sphere count as one each, whereas the mutually stabilizing supporting point contacts contribute as half contact to avoid double counting. Consequently, the {\it apparent} number of supporting point contacts associated with a particle in state $\alpha$ can be calculated as $3-\alpha /2$. The mean coordination number $<Z>$ can then be obtained as

\begin{equation}
<Z> = 2 \sum_{\alpha=0}^{3} {(3-\frac{\alpha}{2}) p(\alpha)}.
\label{1}
\end{equation} It is easy to see that in the simplest case of a sequentially deposited system, where $p(\alpha=0)=1$, Eq. \ref{1} yields the well known value $<Z>=6$.

\subsection{Size distribution}

The size distribution $p(n)$ is defined as the probability that a given particle in the assembly belongs to a bridge of size n, that is

\begin{equation}
p(n) = \frac{nN(n)}{N_{tot}},
\end{equation} where $N(n)$ is the number of bridges consisting of $n$ particles and $ N_{tot}$ is the total number of particles in the system.

In Fig. 4 we can see that different shaking amplitudes do not promote a redistribution of arch sizes in the system. We also find that the total number of bridges does not vary very much as we increase the volume fraction of the system. We have shown \cite{Pugnaloni} that over a large range of bridge sizes the size distribution can be fitted by a power law [$p(n) \propto n^{- \alpha}$] with exponent $\alpha=1.00 \pm 0.03$.

\subsection{\label{maxis}Main axis and moments of inertia}

The principal moment of inertia of the bridges can be obtained by calculating the tensor of inertia of the bridge and then diagonalizing the corresponding matrix. However, a more interesting measure is the moment of inertia around a characteristic axis of the bridge. There is no single way of choosing such a characteristic axis but, in principle, any definition that provides a measure for the orientation of the bridge with respect to the laboratory system of reference would be appropriate.

In two dimensions, the straight line that joins the centres of the only two base-particles (or a perpendicular vector to it, see Fig. 1) would be a good measure of the orientation of a bridge. In three dimensions, unfortunately, a bridge can rest on top of several base-particles (at least three) and therefore there is not a single way of choosing a particular direction except for the one-particle bridges.

We define the main axis of a three-dimensional bridge to have the direction of the vector that results from the average of the normal vectors to the triangles formed by the centres of all the possible combinations of three base-particles of the bridge. This average is weighted by the area of the corresponding triangles. In addition, the main axis passes through the centre of mass of the bridge (see Fig. 2).

Once we have defined the main axis of a bridge of $n$ particles, we can calculate the moment of inertia $M$ around this axis as

\begin{equation}
M=\frac{1}{n}\sum_{i=1}^{n}{(\textbf{r}_{i}\times \textbf{u})^{2}},
\end{equation}
where $\textbf{r}_{i}$ is the position of particle $i$ in the bridge and $\textbf{u}$ is a unit vector with the direction of the main axis that passes through the centre of mass of the bridge. Both $\textbf{r}_{i}$ and $\textbf{u}$ are measured with the centre of mass of the bridge as the origin of the reference system.

The moments of inertia are represented as a function of the bridge size in Fig. 5. Like the size distribution function, the moments of inertia are virtually independent of the volume fraction of the packing of spheres. This is to be expected, as the possible configurations for a bridge of a particular size are not very different in regard to the mass distribution within the bridge. Indeed, we have shown before \cite{Pugnaloni} that a large proportion of the bridges have a string-like structure (a sequence of mutually stable particles supported by two base-particles at each end plus an extra base-particle for each middle bridge-particle). As a result, the mass distribution of a majority of the bridges corresponds to a more or less curved string yielding similar moments of inertia. For large bridges, the data follow a power law with exponent 1.9.

\subsection{Sharpness}

The sharpness of a bridge can be defined as the ratio between a characteristic vertical dimension and a characteristic horizontal dimension of the arch. However, as most of the arches in the interior of an assembly of grains do not lie in a horizontal position, it is more sensible to use the main axis of the bridge as the effective ``vertical" direction.

We define the sharpness of a bridge as the ratio between the maximum projection on the main axis of the distances between all the pairs of particles in the bridge and the square root of the moment of inertia around the main axis. This is, of course, a non-dimensional quantity.

As with the previous structural descriptors, no important discrepancies can be seen in the sharpness of the bridges for different packing fractions (see Fig. 6). An exception can be made for the two-particle bridges, which show a slight increase in sharpness with increasing volume fractions. A power law with exponent -1/2 seems to be followed by large bridges. This decrease in sharpness for increasing bridge sizes implies that bridges grow in height more slowly than in horizontal expansion.

\subsection{Orientation}

In section \ref{maxis} we defined the characteristic axis for a bridge. Therefore, a measure of the orientation of a bridge can be obtained by using the orientation of its main axis. In this case we use the angle $\theta$ between the main axis of the bridge and the vertical direction as an indicator of the ``horizontality" of the bridge. Then, a completely horizontal bridge (i.e. with a ``flat'' base) corresponds to $\theta=0$. 

In Fig. 7 we show the distribution of orientations $\theta$ for different volume fractions (we only include ``real" bridges, i.e. no one-particle bridges are considered). It is clear that most of the bridges have a deviation from the horizontal position of about 20 degrees. However, the actual distribution of the orientation is rather sensitive to the volume fraction of the system: the higher the volume fraction the more horizontal the bridges. So far, this is the most density dependent descriptor we have encountered for the bridges. It is as if a sphere deposit increases its compaction during small amplitude shaking by moving its bridges into a more horizontal orientation rather than changing the actual size distribution or sharpness of the bridges. This was previously suggested by Mehta and Barker \cite{Mehta} based on the lack of correlation between variations in the mean coordination number and the volume fraction of the deposits.

\subsection{Spatial distribution}

The distribution of bridges across the system is also an interesting feature. We have measured the number of bridges per unit volume as a function of the height from the bottom of the container. A bridge is considered to be in a particular volume element if its centre of mass is within that volume element.
Fig. 8 shows that loose packings tend to present a fairly homogeneous distribution of the bridges across the container whereas more compacted systems present a larger number of bridges at the bottom of the container.

We have seen above (Fig. 7) that the overall orientation of the bridges depends on the volume fraction of the system. In Fig. 9 we can see that for highly compacted deposits the bridges are more horizontal at the bottom of the system than anywhere else. However, for loose packings the mean orientation of the bridges is constant throughout the deposit. This stresses the idea that compaction and de-compaction though shaking is directly linked to the reorientation of the arches within the deposit. At the bottom of a highly compact system, the volume fraction is normally higher and so the bridges are more horizontal. On the other hand, loose systems have a constant volume fraction across the deposit presenting then a constant orientation of the arches irrespectively of the z-position in the container.

The results shown in Figs. 8 and 9 are only preliminary evidences due to the small system sizes simulated. Error bars are still rather large due to the lack of statistics. Each data point corresponds to an average carried out only within a region roughly equivalent to one-fifth of the total size of the system.

\subsection{Volume of tetrahedra}

For completeness, we consider here the volume of the tetrahedron \cite{Gotoh} formed by the centres of a particle and its three supporting particles (see Fig. 10). Although this is not a specific property of the bridges, it is very sensitive to the volume fraction of the deposit.

For a sequentially deposited system, where no arches exist, the volume of the tetrahedron formed by a particle and its supporting particles is relatively well defined. As we can see in Fig. 10, the volume distribution for a sequential system presents a very sharp peak around $0.14 \sigma^{3}$. Much bigger volumes cannot be obtained because there is a maximum possible value of $\sigma^{3}/6$, corresponding to base particles located on the vertexes of a horizontal equilateral triangle of side length $\sqrt{2}$. However, small volumes can be obtained in principle by separating the supporting particles so that the three contact points are near their equators. It is clear from Fig. 10 that these configurations are highly improbable to occur in sequentially deposited granular packings.

Highly contrasting with the previous scenario, a shaken deposit presents an almost uniform distribution of volume for the basic tetrahedra. Nevertheless, a memory of the characteristic peak, which is present in sequentially deposited systems, remains clear. This broad distribution of the tetrahedron volumes is to be expected from mutually stabilized particles, as they tend to have their contact points closer to the equators (see Fig. 2). This is in agreement with previous studies \cite{Nolan} of the projection of the centre-to-centre vectors of a sphere and its supporting spheres along the Cartesian axis. Sequentially deposited systems present vertically biased projections whereas non-sequentially rearranged structures show horizontally biased projections.

It is worth mentioning that the use of ``the most probable tetrahedron'' \cite{Gotoh} as a way to estimate the volume fraction might be inadequate in the case of shaken deposits because of the broad distribution of tetrahedral shapes they present. However, sequentially deposited packings should be fairly well described using the most probable shape of the characteristic thetrahedra.

An interesting feature that can also be appreciated in Fig. 10, is that as bridges become increasingly ``horizontal" (for increasing volume fractions), the tetrahedron formed by a particle and its base becomes increasingly sharper leading to larger volumes. This seems to be in contradiction with the fact that the system is increasing its volume fraction. However, we have to recall that the volume of the supporting tetrahedra are not related to the actual volume fraction of the packing in a simple way \cite{Gotoh}.
 \section{Conclusions}

We have considered here several structural descriptors for the arches formed by inelastic hard spheres in non-sequentially deposited systems. We find that descriptors related to the shape and size of the arches ---such as the size distribution, moments of inertia and sharpness--- show no strong dependency on the packing fraction of the deposits. Although these are greatly idealized granular systems (mono-sized hard spheres), we can speculate that these descriptors might have some universal qualities.

On the other hand, the orientation and spatial distribution of the bridges appear to be very sensitive to the packing fraction of the system. Previously, Mehta and Barker \cite{Mehta} suggested that, since changes in volume fraction in these systems are not due to changes in the mean coordination number, it is the shape of bridges rather than their number that determines the compaction. These new results suggest that changes in the volume of a granular packing are driven by the reorientation of the bridges inside it ---the more horizontal the bridges, the more dense the deposit--- rather than by a change in bridge sizes or shapes. 

Finally, the shape of the basic tetrahedra formed by a sphere and its supporting spheres shows a qualitative change when sequentially deposited systems are compared against non-sequentially reorganized deposits.

It is worth mentioning that the lateral extension of the systems simulated in this work is rather small ($6\sigma \times 6\sigma$). Consequently, the statistics for large bridges ---larger than 20 spheres or so--- can be largely misleading. Nevertheless, the actual value of those structural descriptors that are averaged over all bridges ---such as the orientation distribution of the bridges--- is indeed reliable since it is dictated by the large population of small bridges in the system. The orientation of the bridges depends on the bridge size but the overall distribution is close to the distribution for small bridges. In Fig. 11, we show an example of the orientation distribution for bridges of sizes 2, 3, 4, 8 and 9 along with the overall distribution from Fig. 7. Bridges of large size are more horizontal than small bridges. Further investigations on these types of orientation distributions discriminated by bridge size can be found in Ref. \cite{Mehta3}. We have also carried out size-weighted averages for the bridges. Although the distributions are slightly different, the same discrepancies between systems at different volume fractions have been observed.

More detailed studies on the properties of the bridges might lead to a better understanding of the relationship between the shaking amplitude and the packing fraction of shaken granular packings at their steady state, the jamming of flowing grains through small openings, and the force distribution across the packing contact network. Moreover, the correlation between arches found in consecutive shaking cycles could shed light on the microscopic mechanisms that drive the ``irreversible" and ``reversible" branches of the density of granular materials subject to varying shaking amplitudes.
 
\bigskip
 \bigskip \bigskip \bigskip \bigskip

{\bf ACKNOWLEDGMENTS}

LAP thanks support from Fundaci\'{o}n Antorchas (Argentina) and the British Council, and the Spanish Ministry of Science and Technology (Project BFM2002-00414). GCB acknowledges support from the Biotechnology and Biological Science Research Council, UK; and many valuable discussions with Dr. Anita Mehta.

\begin{center}
{\bf Figure Captions}
\end{center}

{\bf Figure 1:} Example of bridges in two dimensions. A) A one-particle bridge and its two base-particles. B) A three-particle bridge and its two base-particles. The main axis of the bridge is indicated along with the orientation angle.
 
{\bf Figure 2:} Example of bridges in three dimensions found in our simulated deposits. A) A two-particle bridge with four base-particles (two-particle bridges with three base-particles also exist although they are very rare). B) A four-particle bridge with six base-particles. The main axis of the bridges are indicated along with the orientation angle.

{\bf Figure 3:} Probability $p(\alpha)$ of finding a particle with state of stability $\alpha$ for different volume fractions. For a sequentially deposited set of spheres, no mutual stabilization is possible, i.e. $p(0)=1$. Error bars correspond to the standard deviation.

{\bf Figure 4:} Log-log plot of the bridge size distribution $p(n)$ for different volume fractions. The solid line corresponds to a power law with exponent -1.0.

{\bf Figure 5:} The moment of inertia of the bridges around their main axis against the bridge size for different volume fractions. The solid line corresponds to a power law with exponent 1.9.

{\bf Figure 6:} Sharpness of the bridges as a function of their size for different volume fractions. See text for details. The solid line corresponds to a power law with exponent -1/2.

{\bf Figure 7:} Orientation distribution of the bridges for different volume fractions. Horizontal bridges correspond to $\theta=0$. Bridges of size one are not taken into account.

{\bf Figure 8:} Vertical distribution of the bridges for different volume fractions. One-particle bridges are not considered in this distribution. The error bars correspond to the standard deviation and are similar in all the series of data although displayed in only one of them for clarity.

{\bf Figure 9:} Mean orientation of the bridges as a function of their height in the packing for different volume fractions. One-particle bridges are not taken into account. The error bars correspond to the standard deviation and are similar in all the series of data. 

{\bf Figure 10:} Volume distribution of the basic tetrahedra in the sphere deposit for different volume fractions. The result for a sequentially deposited system is included for comparison.

{\bf Figure 11:} Orientation distribution of n-bridges for $\phi=0.56$. Bridges of sizes 2, 3, 4, 8 and 9 are considered. The solid line is the corresponding overall orientation distribution (see Fig. 7).

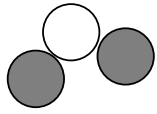
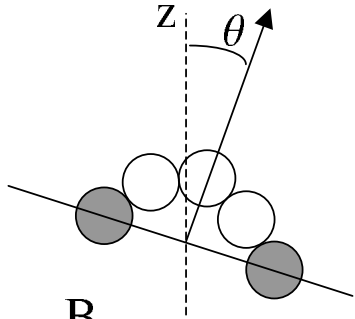
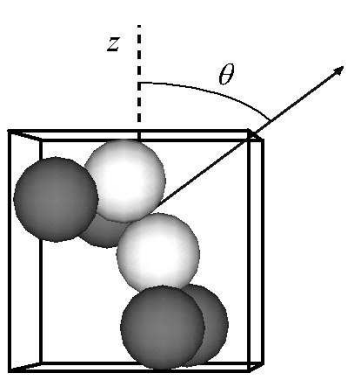
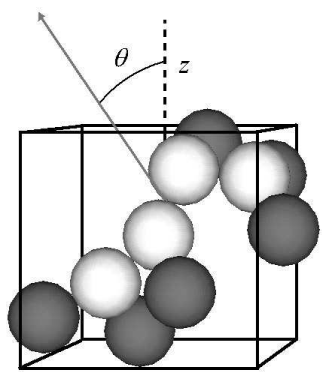
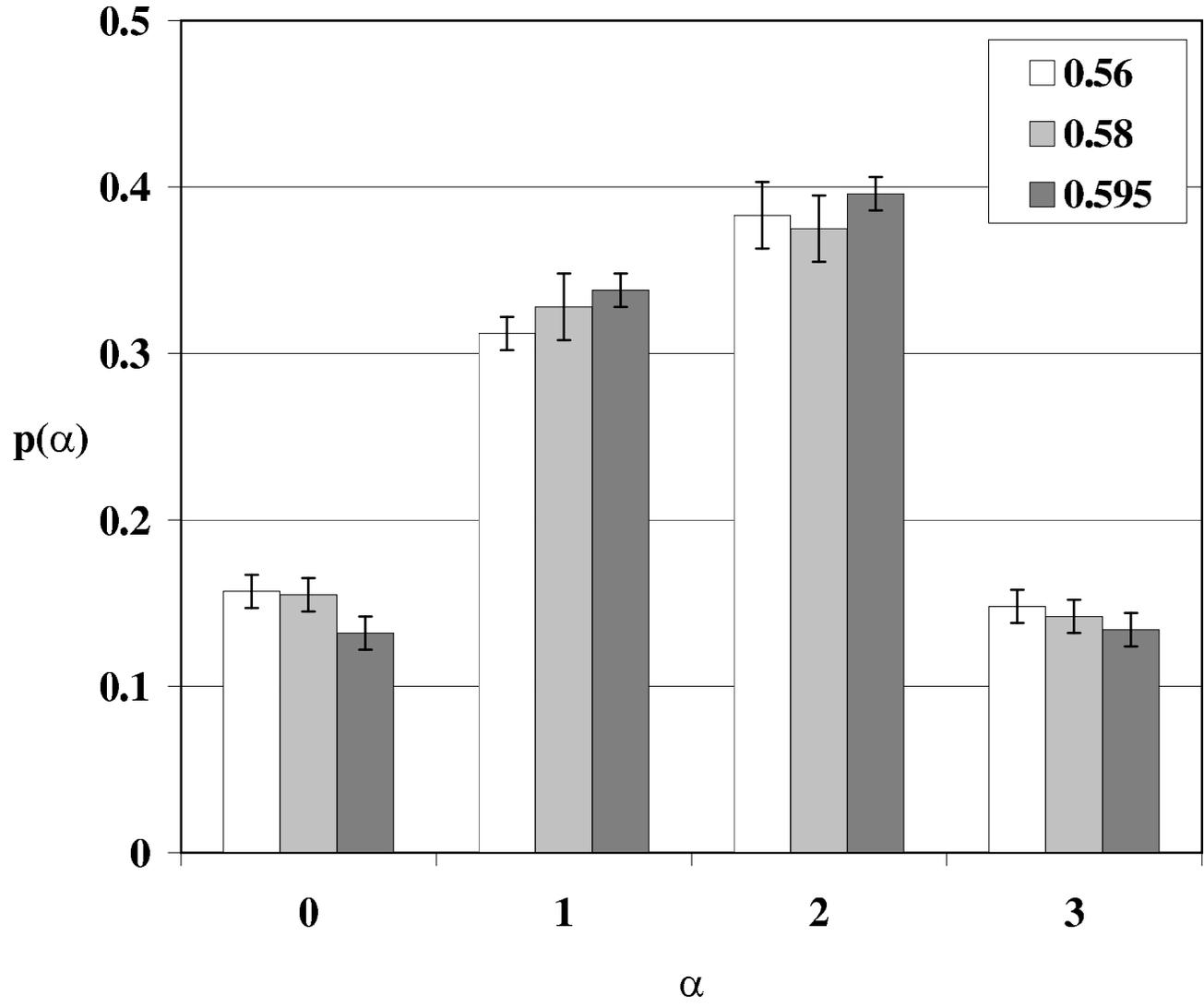
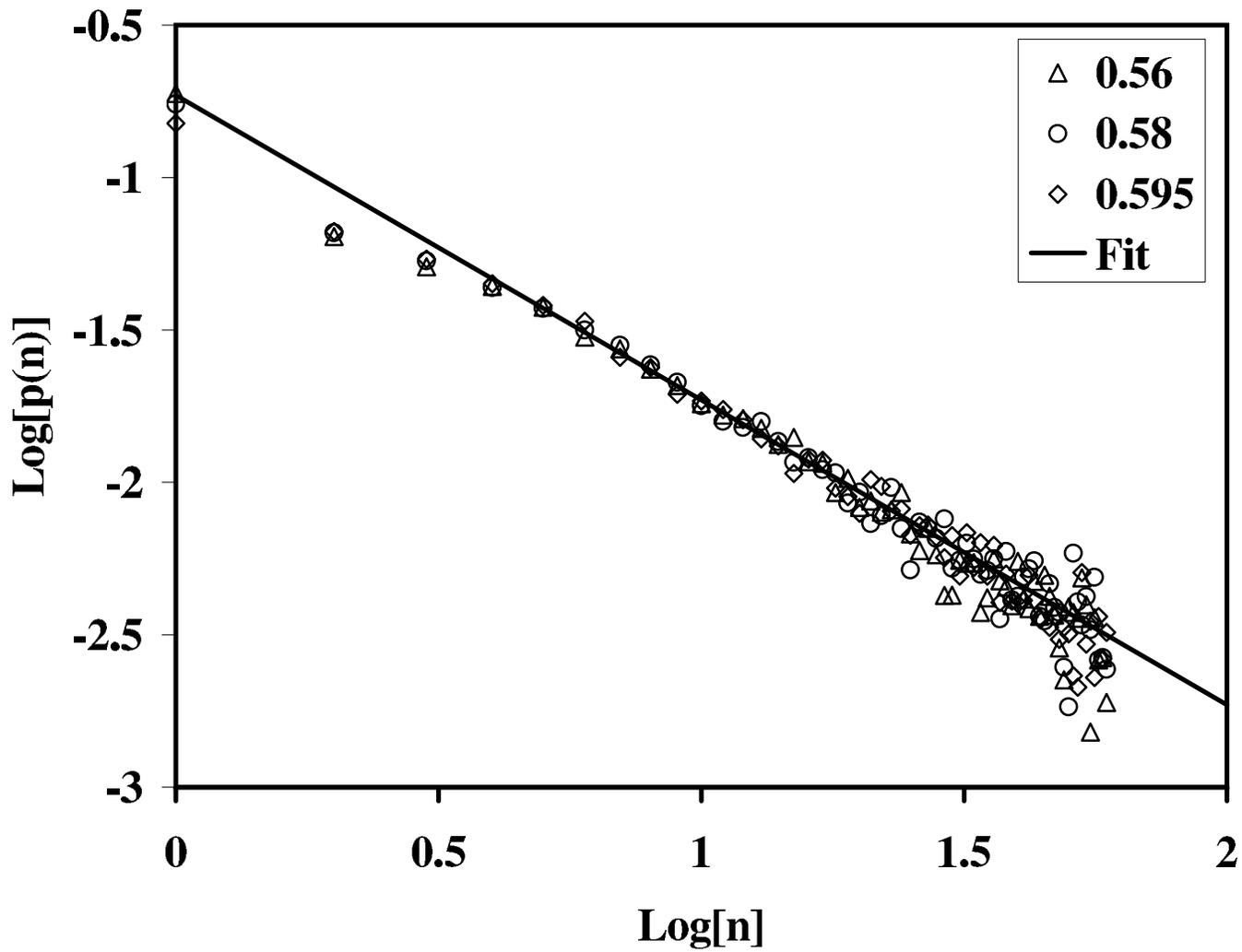
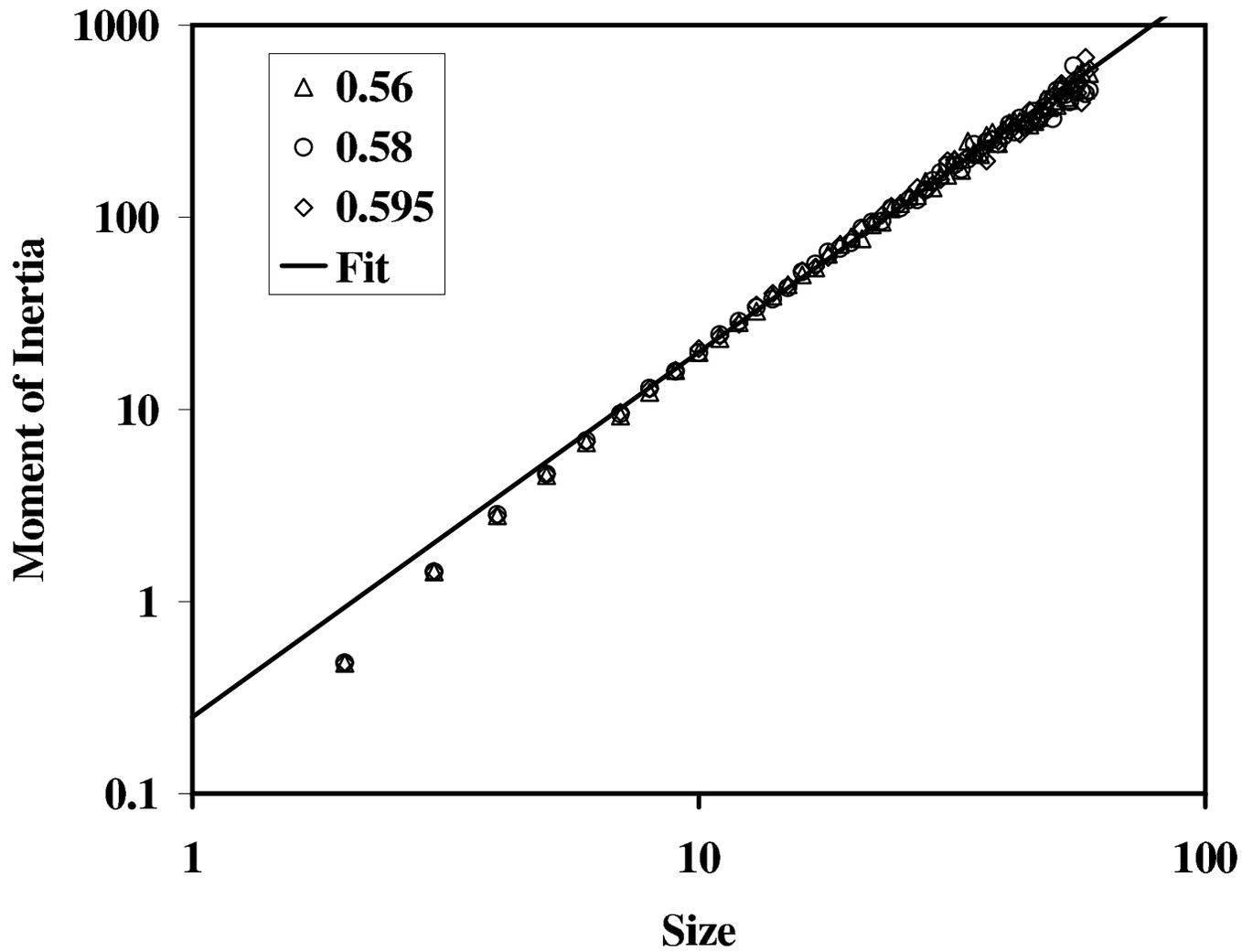
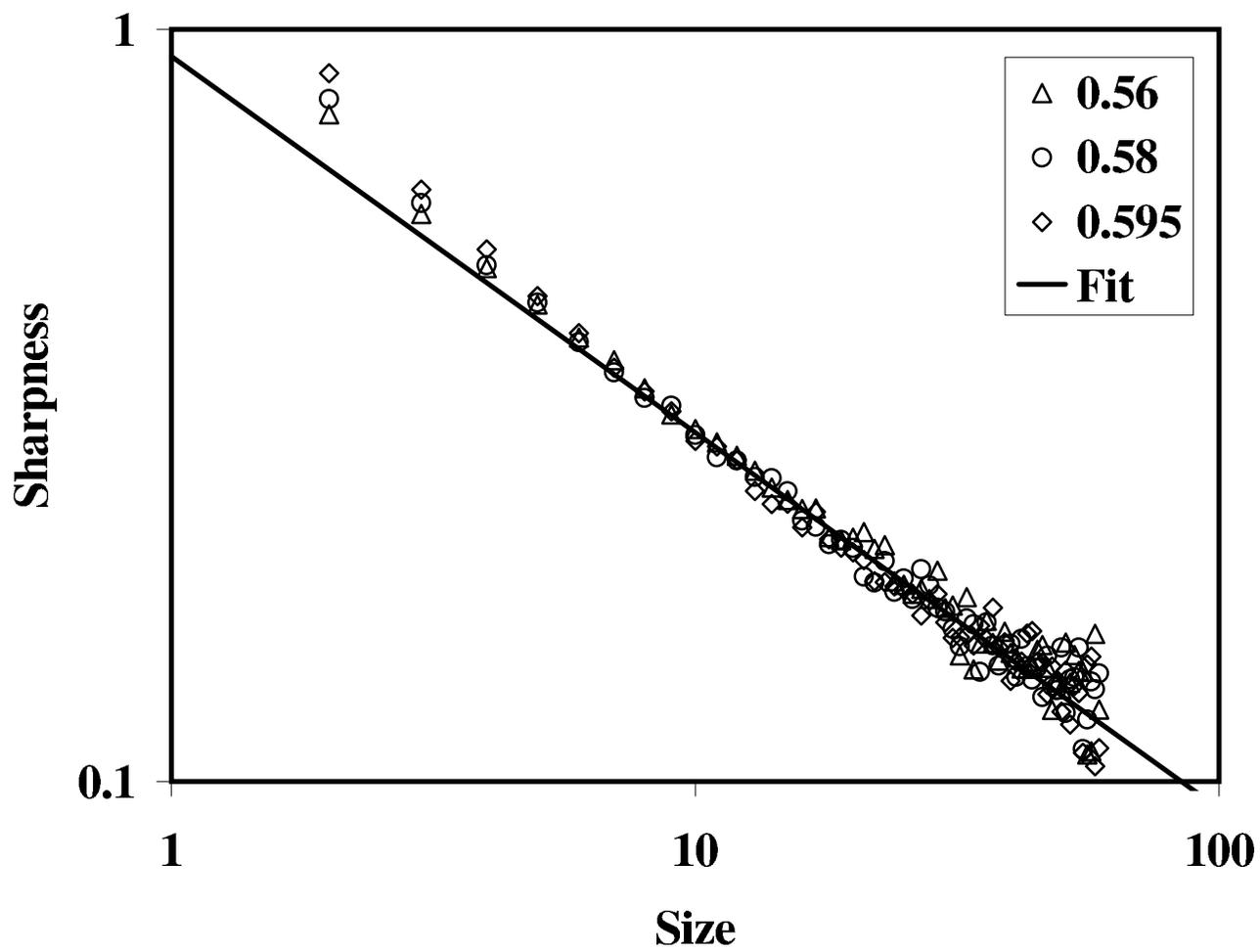
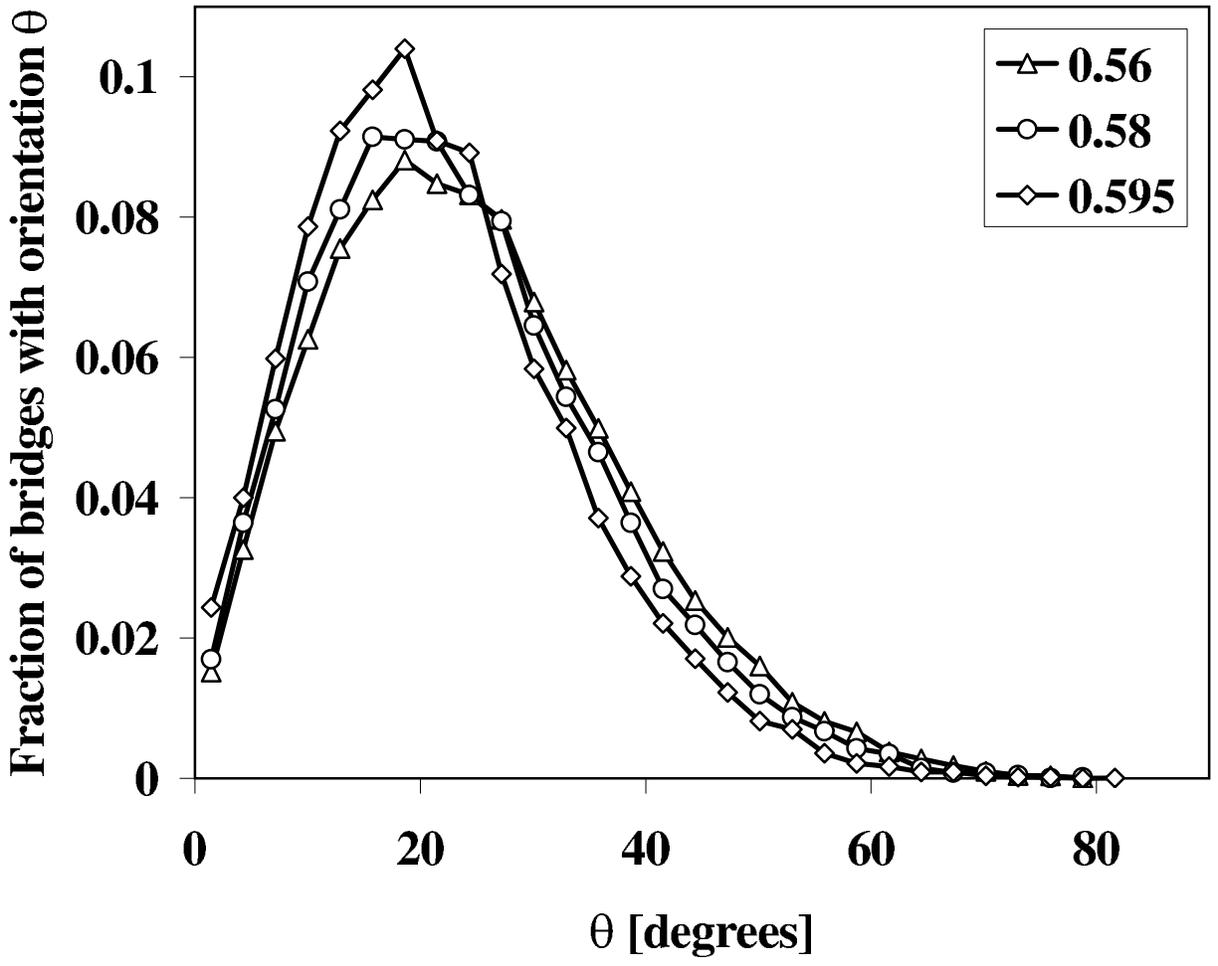
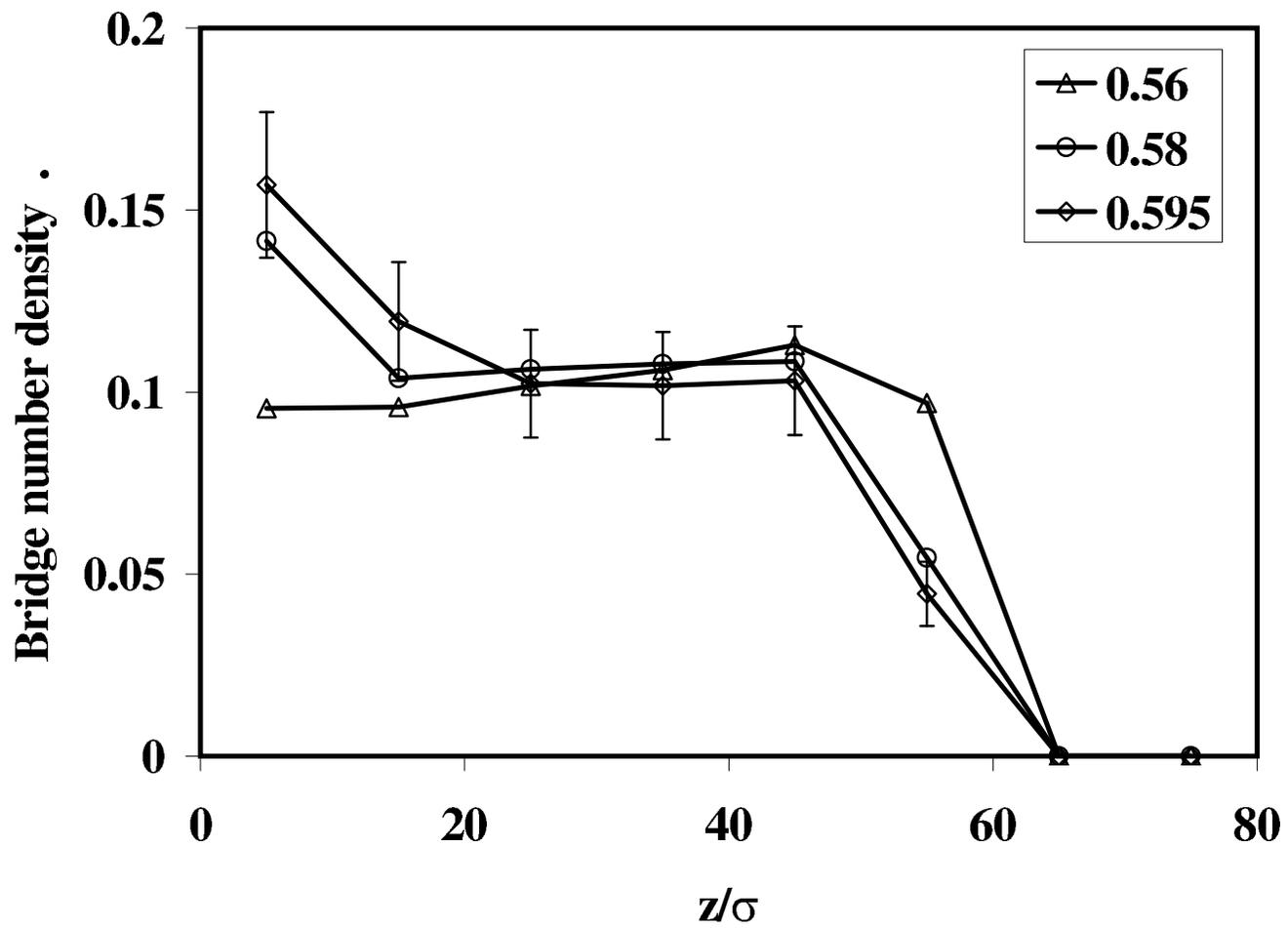
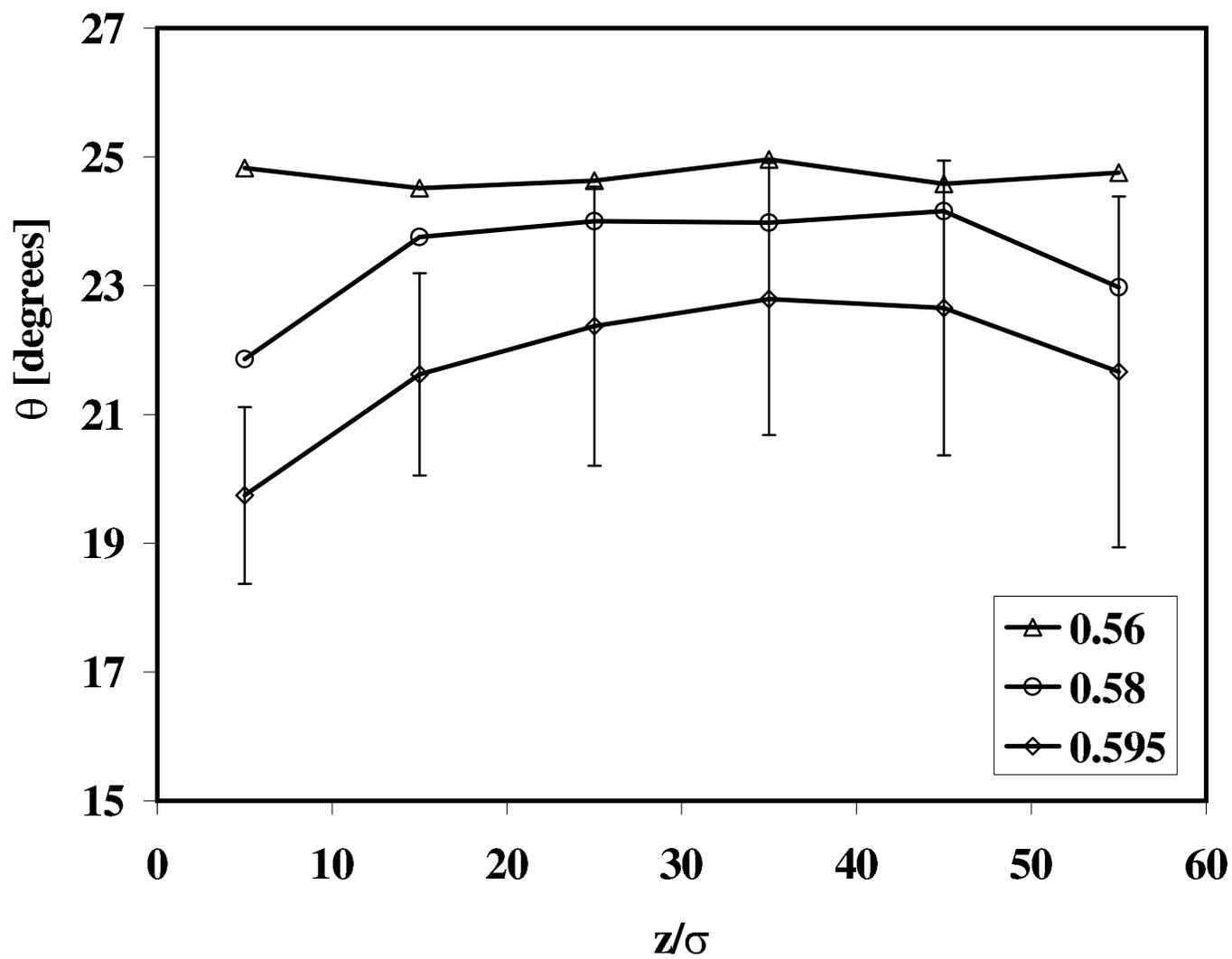
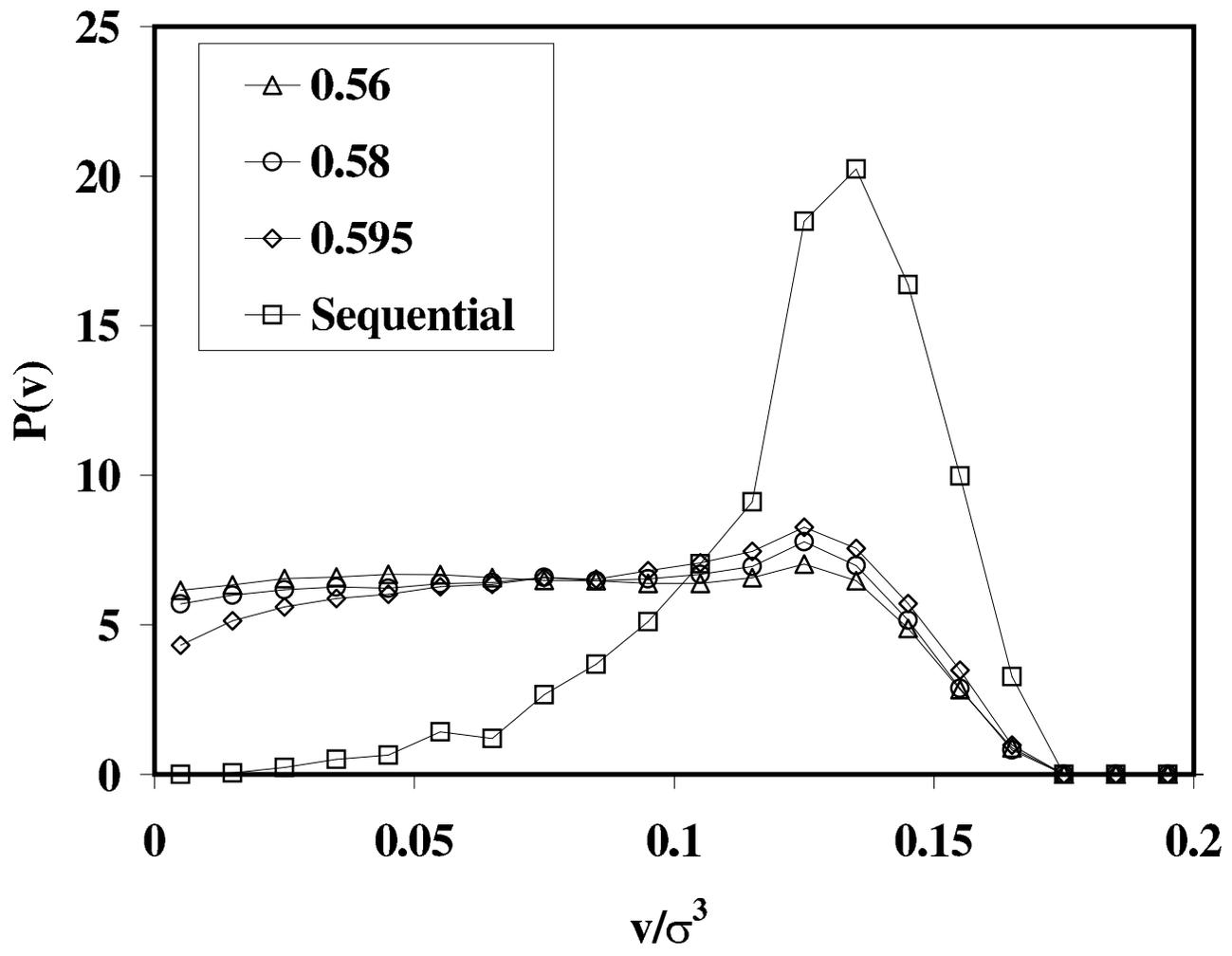
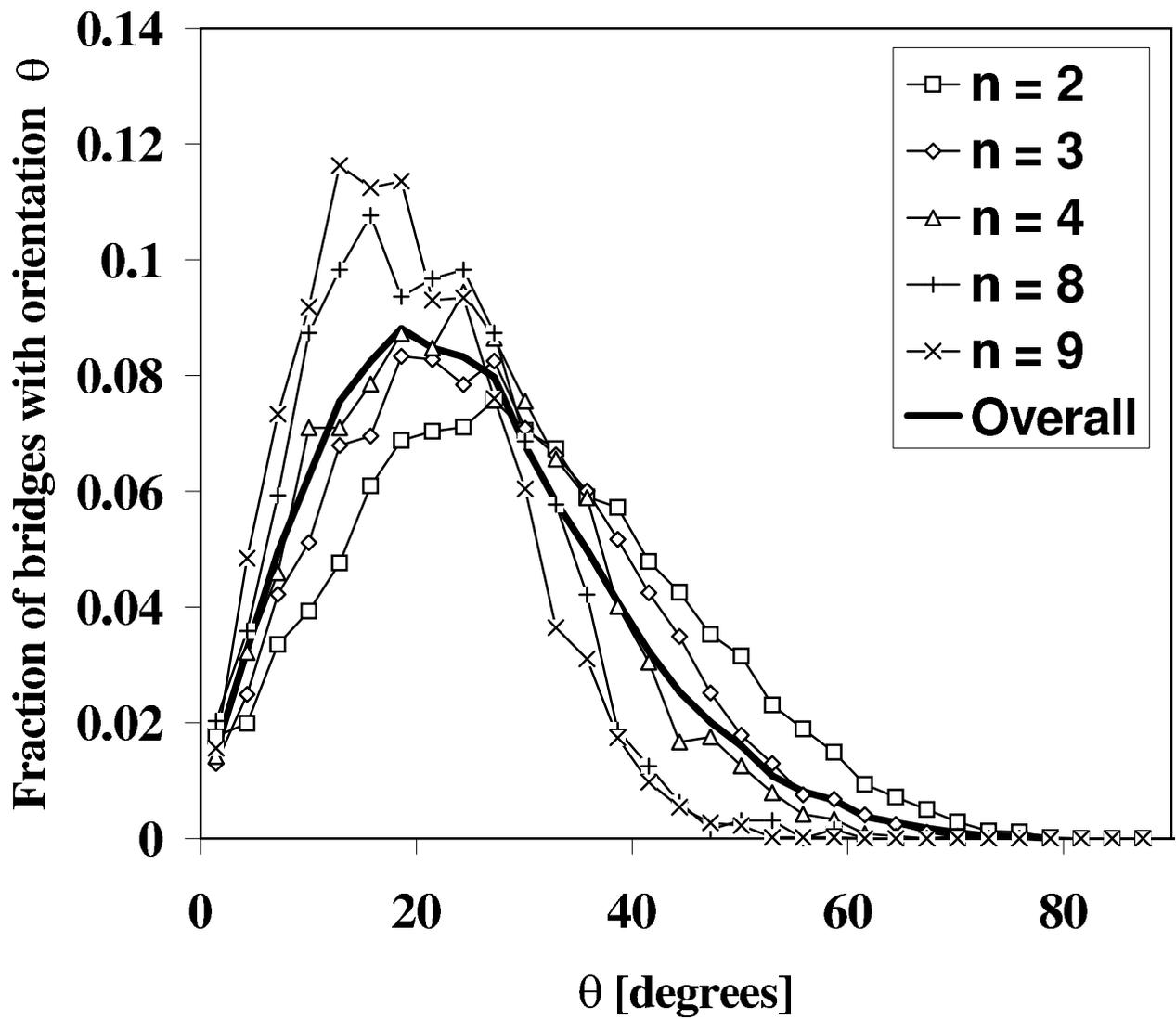
\end{document}